\documentclass{emulateapj}

\newcommand{\kms}{\mbox{km s$^{-1}~$}} 
\newcommand{\kmse}{\mbox{km s$^{-1}$}} 
 
\newcommand{\msun}{M$_{\odot}~$} 
\newcommand{\msune}{M$_{\odot}$} 
\newcommand{\vlsr}{$V_{\rm LSR}~$}
\newcommand{\vlsre}{$V_{\rm LSR}$}

\newcommand{\dgr}{$^{\circ}~$}

\newcommand{\hi}{\ion{H}{1} }
\newcommand{\hie}{\ion{H}{1}}

\newcommand{\nhie}{$N_{\rm H \small{I}}$}
\newcommand{\et}{et al.}

\begin{document}

\title{Evidence for an Interaction in the nearest \\
Starbursting Dwarf Irregular Galaxy IC~10}

\shorttitle{IC~10 \hi Stream}
\shortauthors{NIDEVER ET AL.}

\author{David L. Nidever\altaffilmark{1,2},
Trisha Ashley\altaffilmark{3},
Colin T. Slater\altaffilmark{1},
J\"urgen Ott\altaffilmark{4},
Megan Johnson\altaffilmark{5}, \\
Eric F. Bell\altaffilmark{1},
Sne\v{z}ana Stanimirovi{\'c}\altaffilmark{6},
Mary Putman\altaffilmark{7},
Steven R. Majewski\altaffilmark{2}, \\
Caroline E. Simpson\altaffilmark{3},
Eva J\"utte\altaffilmark{8},
Tom A. Oosterloo\altaffilmark{9,10}, and
W. Butler Burton\altaffilmark{11}
}

\altaffiltext{1}{Department of Astronomy, University of Michigan,
Ann Arbor, MI, 48109, USA (dnidever@umich.edu)}

\altaffiltext{2}{Department of Astronomy, University of Virginia,
Charlottesville, VA, 22904-4325, USA}

\altaffiltext{3}{Department of Physics, Florida International University, Miami, FL, 33199, USA}

\altaffiltext{4}{National Radio Astronomy Observatory, Socorro, NM, 87801, USA}

\altaffiltext{5}{National Radio Astronomy Observatory, Green Bank, WV, 24944, USA}

\altaffiltext{6}{Department of Astronomy, University of Wisconsin, Madison, WI, 53706, USA}

\altaffiltext{7}{Department of Astronomy, Columbia University, New York, NY, 10027, USA}

\altaffiltext{8}{Astronomisches Institut der Ruhr-Universit\"at Bochum, Universit\"atsstr. 150, 44801 Bochum, Germany}

\altaffiltext{9}{Netherlands Institute for Radio Astronomy (ASTRON), Postbus 2, 7990 AA Dwingeloo, The Netherlands}

\altaffiltext{10}{Kapteyn Astronomical Institute, University of Groningen, Postbus 800, 9700 AA Groningen, The Netherlands}

\altaffiltext{11}{National Radio Astronomy Observatory, Charlottesville, VA, 22903, USA}

\begin{abstract}
Using deep 21-cm \hi data from the Green Bank Telescope
we have detected an $\gtrsim$18.3 kpc--long gaseous extension associated with the starbursting dwarf
galaxy IC~10.  The newly-found feature stretches 1.3\dgr to the northwest and has a large radial velocity
gradient reaching to $\sim$65 \kms lower than the IC~10 systemic velocity.  A region of higher
column density at the end of the extension that possesses a coherent velocity gradient ($\sim$10 \kms across
$\sim$26\arcmin) transverse to the extension suggests rotation and may be a satellite galaxy of IC~10.
The \hi mass of IC~10 is 9.5$\times$10$^7$ ($d$/805 kpc)$^2$ \msun and the mass of the
new extension is 7.1$\times$10$^5$ ($d$/805 kpc)$^2$ \msune.
An IC~10--M31 orbit using known radial velocity and proper motion values for IC~10 show
that the \hi extension is inconsistent with the trailing portion of the orbit so that an M31--tidal
or ram pressure origin seems unlikely.  We argue that the most plausible explanation for the new feature is
that it is the result of a recent interaction (and possible late merger) with another dwarf galaxy.
This interaction could not only have triggered the origin of the recent starburst in IC~10, but could
also explain the existence of previously-found counter-rotating \hi gas in the periphery
of the IC~10 which was interpreted as originating from primordial gas infall.
\end{abstract}

\keywords{galaxies: dwarf --- galaxies: individual (IC 10) --- galaxies: interactions ---
galaxies: kinematics and dynamics --- galaxies: starburst --- Local Group}

\section{Introduction}
\label{sec:intro}

IC~10 is a blue compact dwarf \citep[BCD;][]{richer01} galaxy in the Local Group (LG) and is the closest
known starburst galaxy.  It is one of the more distant satellite galaxies in the M31 group
($d_{\rm M31}$$\sim$250 kpc) and lies in the thin plane of M31 corotating satellite galaxies
\citep{Ibata13,Tully13}.  IC~10 is believed to be currently isolated from other known galaxies \citep{hunter04},
but studies of its youngest stars show that the starburst is only $\sim$10 Myr old \citep{Massey07}.

What regulates star formation in BCDs, and dwarf galaxies in general, is not well understood. 
BCDs have high star formation rates that give rise to bluer colors
than for other dwarf galaxies \citep{kunth95} and were originally singled out for their compact stellar
appearance on photographic plates \citep{zwicky66}.  BCDs cannot long sustain a starburst; they should
deplete their gas reservoirs in $\thicksim$$10^9$ years without additional gas accretion \citep{gil05}.
Thus, it is likely that something has recently triggered this burst of star formation \citep{schulte01, crone02}.

It has often been speculated that most BCDs are the result of a recent dwarf-dwarf merger or a gravitational
interaction \citep{noeske01, Pustilnik01b, bekki08}, but studies show that there are still many BCDs found
with no nearby companions \citep{vanzee01, zitrin09, koulouridis13}, making a merger or recent
interaction unlikely. In cases such as these other possible triggers have been adopted, for example:
faint companions, past mergers, intergalactic gas accretion, dark matter satellites, and gas sloshing
about in dark matter
potentials \citep{helmi12, Simpson11, noeske01}.  IC~10's proximity \citep[805 kpc;][]{Sanna08} and 
apparent isolation make it an excellent candidate for studying star formation and alternative triggers in
starburst dwarf galaxies. 

Previous atomic hydrogen (\hie) studies of IC~10 found a wide \hi envelope \citep{Huchtmeier79}, an inner
rotating disk, and counter-rotating gas in the periphery often thought to be caused by infalling primordial
gas \citep[e.g.,][]{Shostak89,Wilcots98}.  The inner region of IC~10 contains \hi holes and shells thought
to be shaped by stellar winds \citep{Wilcots98} from the burst of star formation in IC~10. 

\citet{Wilcots98} analyzed high resolution \hi data of IC~10 and concluded that IC~10 could be the result of
an interaction or merger, but that IC~10's gas is so chaotic that it is more likely a galaxy still in formation,
accreting cosmological gas around it. Still, the possibility that IC~10 is the result of a merger or interaction
could not be ruled out. 

If an interaction or merger has recently occurred in IC~10, then there should be signatures of that event in the \hie,
such as tidal tails or bridges \citep{toomre72}, or components of a previously unknown companion, that may be
too tenuous or large to have been seen in previous studies.  Single dish \hi observations of IC~10 by
\citet{Huchtmeier79} have shown that the \hi pool of IC~10 extends to seven times its optical diameter at
a column density of $3.6\times10^{18}\ \rm{atoms}\ \rm{cm}^{-2}$ with no obvious signatures of a tidal
interaction or merger. However, signatures of environmental influences may be hidden at sensitivities as low as
$\sim$$10^{18}\ \rm{atoms}\ \rm{cm}^{-2}$ \citep[e.g.,][]{Johnson13}.

In this Letter, we present new results from the combination of two Green Bank Telescope surveys
in the vicinity of IC~10 that reveal a long gaseous northern extension.
In Section \ref{sec:red}, we present the observations and their reduction.  In Section \ref{sec:results},
we give results of our analysis of these data, and, finally, in Section \ref{sec:origin} we discuss the
origin of the newly-found \hi extension and its significance.

\begin{deluxetable}{lc}
\tablecaption{Properties of IC~10 and new \hi Extension}
\tablewidth{0pt}
\tablehead{
\colhead{Parameter} & \colhead{Value}
}
\startdata
\underline{GBT Observations:} & \\
Resolution & 9.1\arcmin \\
RMS noise & $\sim$21 mK per channel \\
3$\sigma$ sensitivity over 20 \kmse & $\sim$6.5$\times$10$^{17}$ atoms cm$^{-2}$ \\
& \\
\underline{IC~10 galaxy:} & \\
Coordinates (J2000) &  $\alpha$=00:20:23.16, $\delta$=+59:17:34 \\
Coordinates (Galactic) & $l$=118.97\degr, $b$=$-$3.334\degr \\
Distance & 805 kpc \\
$M_{\rm H \small{I}}$ & 9.5$\times$10$^7$ \msun \\
& \\
\underline{New northern extension:} & \\
Length & $\sim$1.3\degr, $\sim$18.3 kpc \\
Width &  $\sim$0.37\degr, $\sim$5.2 kpc \\
Orientation &  $\sim$25\dgr west of north \\
Velocity offset & $\sim$65 \kms below systemic \\
$<$\nhie$>$ & $\sim$7$\times$10$^{17}$ atoms cm$^{-2}$ \\
$M_{\rm H \small{I}}$ & $\sim$7.1$\times$10$^5$ \msun
\enddata
\end{deluxetable}

\section{Observations And Data Reduction}
\label{sec:red}

Data from two independent Robert C. Byrd Green Bank Telescope (GBT) \hi surveys were combined to produce
the sensitive datacube of IC~10 used for our analysis:
(1) A survey of the tip of the Magellanic Stream (MS) (D.\ Nidever et al., in prep.), and
(2) a survey of a subset of galaxies from the LITTLE THINGS
survey\footnote{https://science.nrao.edu/science/surveys/littlethings} \citep{Hunter12}.

The MS survey used the GBT to conduct a $\sim$300 deg$^2$, 21--cm survey of the MS-tip
and map the MS emission across the Milky Way (MW) midplane (proposal IDs: GBT10B-035
and GBT11B-082).  The results and more details on our MS-tip GBT survey will be presented in a future
paper (D.\ Nidever et al., in prep.).

The MS-tip observations used the ``On-the-Fly'' (OTF) mapping mode (scanning in Right Ascension) to obtain
frequency-switched, 21-cm spectral line data with the Spectrometer backend at a resolution of 0.32 \kmse.  
Each integration (spaced by 4.0$\arcmin$) was covered twice with a total integration time of 18 seconds
on average.  The 435 hours of observing were taken throughout 2010 and 2011.

The MS-tip data reduction was perfomed in a similar manner to that described in \citet[][hereafter N10]{Nidever10}
but with some improvements.  We used the GETFS program in GBTIDL\footnote{http://gbtidl.nrao.edu/} to
obtain calibrated frequency-switched spectra for every position and polarization.  To reduce the noise, the
reference spectra were smoothed with a 16-channel boxcar smoothing box.  Radio frequency interference
(RFI) was automatically detected in both the signal and reference spectra.
RFI is detected as large positive spikes in the average spectrum (as a function of observed frequency, not
velocity) of a scan after a 3-channel median filter has been subtracted to remove real structure.
Channels with RFI were
flagged in the signal spectra but interpolated over (using neighboring ``good'' channels) in the
reference spectra (so as not to contaminate the signal spectra at these channels).  The data were
averaged over five velocity bins (giving a velocity resolution of $\sim$1.6 \kmse) and then our special-purpose
baseline fitting and removal routines (as described in N10) were used.  An additional baseline removal
step was performed to remove some residual structure by fitting a cubic B-spline (with breakpoints
every 50 channels or 80 \kmse) to blocks of 500 integrations with iterative outlier rejection.
The final spectra in a region around IC~10 were resampled onto a grid in Galactic coordinates
at 4$\arcmin$ spacing and multiple passes were combined with exposure time weighting.

\begin{figure}[ht!]
\begin{center}
\includegraphics[angle=0,scale=0.40]{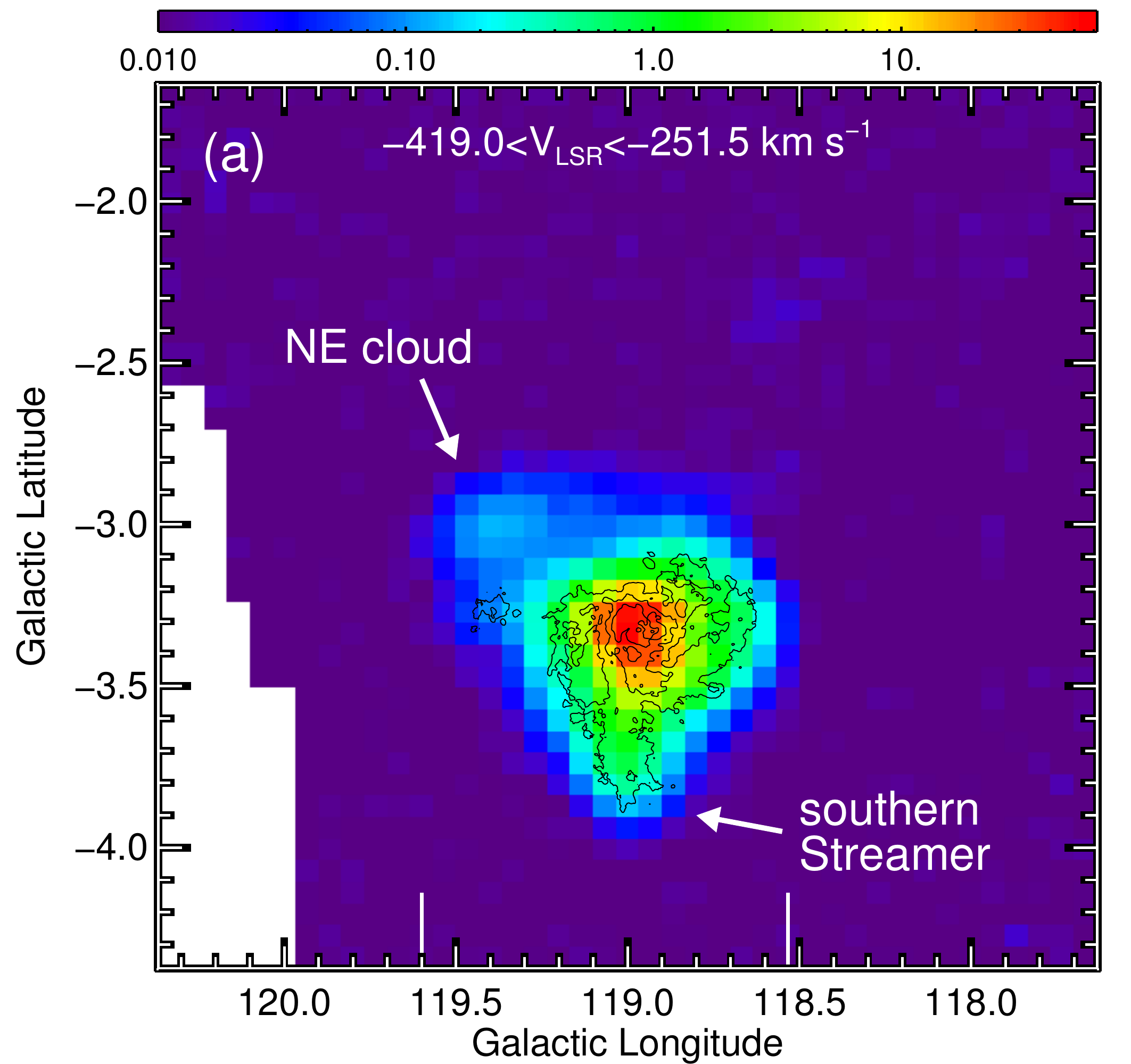}
\includegraphics[angle=0,scale=0.40]{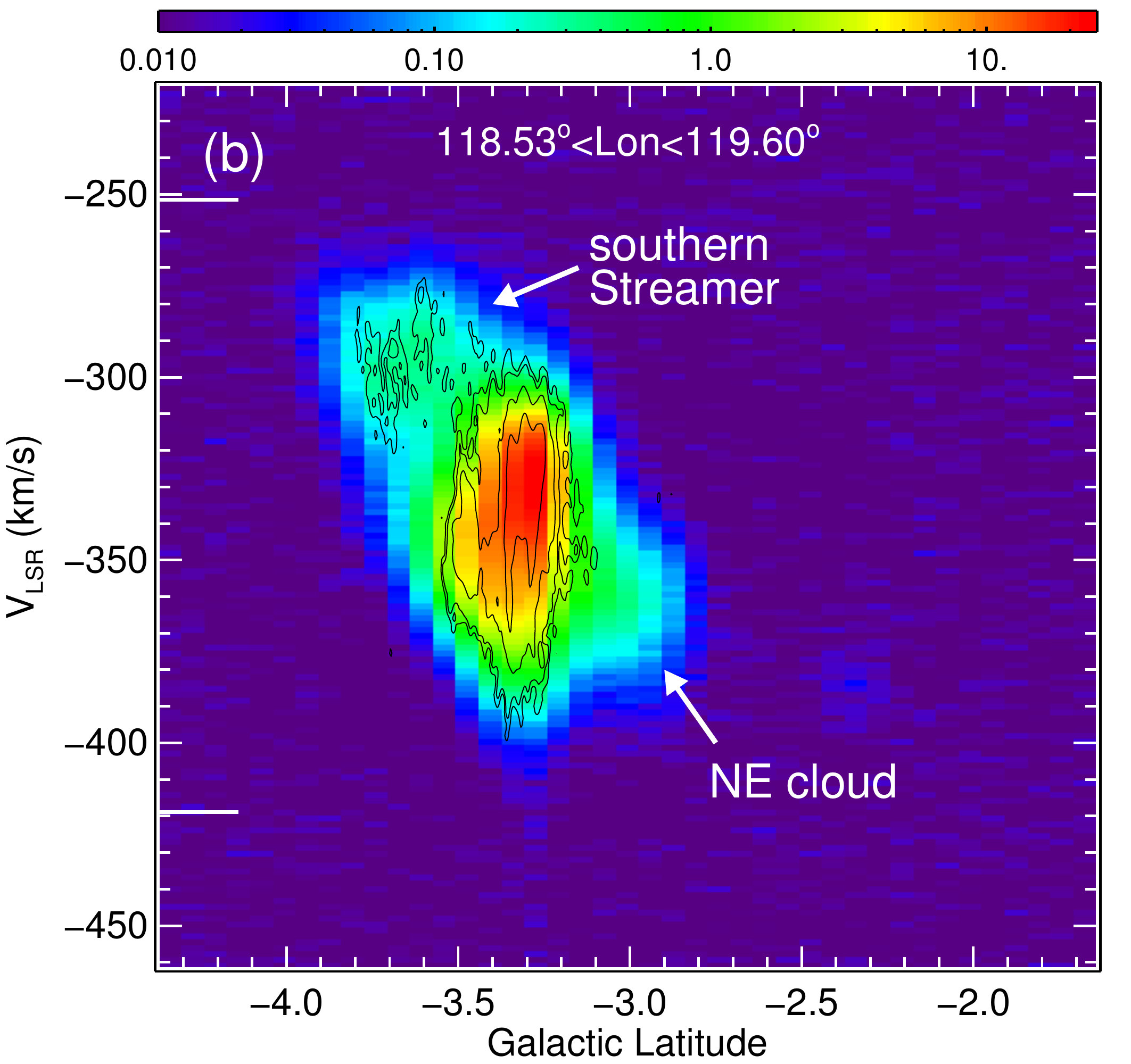}
\end{center}
\caption{{\em (a)} Column density of the entire IC~10 galaxy ($-$419.0$<$\vlsre$<$$-$251.5 \kmse, tick marks
in panel b demarcate the integration range; log(\nhie) in units of 10$^{19}$ cm$^{-2}$).
Contours indicate the column density of the high-resolution WSRT data \citep{Manthey08}.
{\em (b)} Latitude--velocity diagram of IC~10  (118.53$<$$l$$<$119.6\degr, tick marks in panel a demarcate the
integration range).  Contours indicate the distribution of the WSRT data \citep{Manthey08}.}
\label{fig_coldens}
\end{figure}

\begin{figure}[ht!]
\begin{center}
\includegraphics[angle=0,scale=0.40]{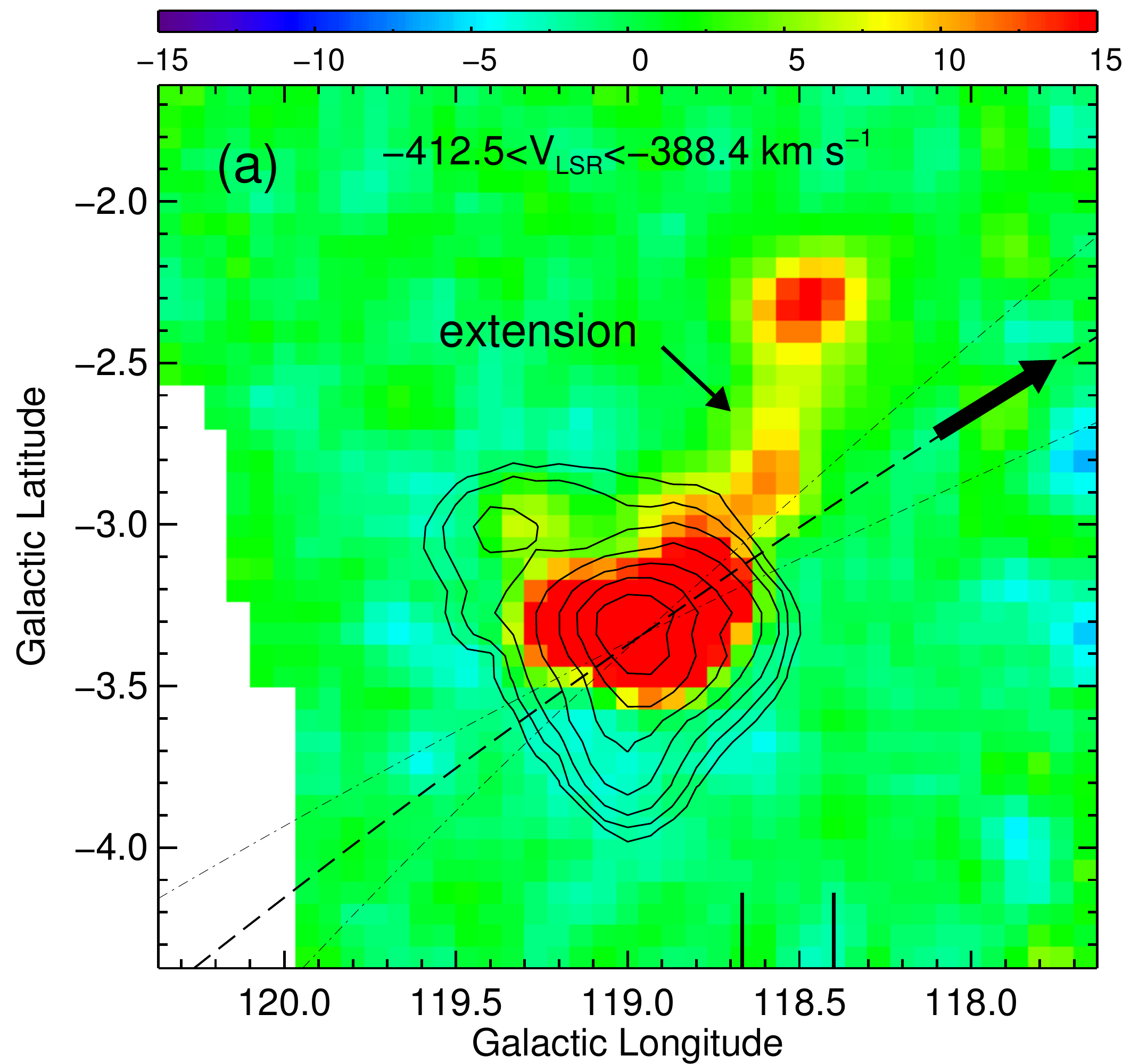}
\includegraphics[angle=0,scale=0.40]{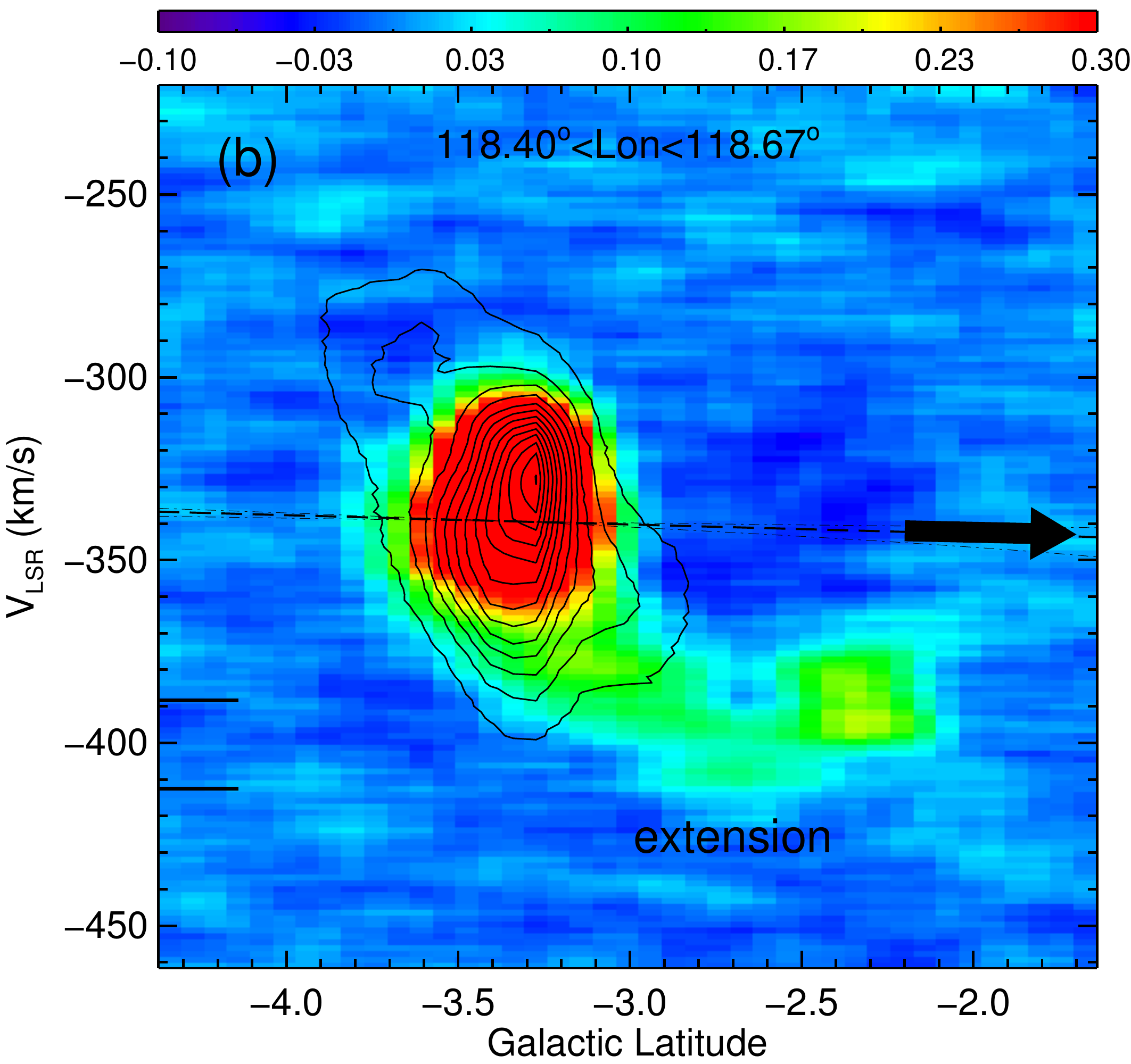}
\end{center}
\caption{{\em (a)} Column density of the new IC~10 extension  ($-$412.5.0$<$\vlsre$<$$-$388.4 \kmse, 
tick marks in panel b demarcate the integration range; smoothed with 12\arcmin$\times$12\arcmin~filter)
in units of 10$^{17}$ cm$^{-2}$.
Contours show the column density from the entire galaxy (as seen in Fig.\ \ref{fig_coldens}a)
at levels of \nhie=0.5--31.5$\times$10$^{19}$ atoms cm$^{-2}$ with steps of 0.3 in log(\nhie).
{\em (b)} Position-velocity diagram of the IC~10 extension (118.4$<$$l$$<$118.67\degr,  tick marks
in panel a demarcate the integration range).  Contours show intensity as seen in Fig.\ \ref{fig_coldens}b.
The dashed line in both panels shows the IC~10 orbit and the dash-dotted lines the 1$\sigma$ uncertainties,
as described in the text.}
\label{fig_coldens_stream}
\end{figure}

Observations for the second survey (proposal ID: GBT13A\_430) were obtained in a manner similar to those by
the MS-tip survey except
that the OTF mapping scanned in Galactic Longitude, the velocity resolution was 0.16 \kmse, and the integrations
were spaced by 3.5$\arcmin$.  The GETFS program was used to obtain calibrated spectra, and RFI were
flagged by hand and interpolated over using non-contaminated neighboring channels.  The data were boxcar
smoothed to a velocity width of 1.6 \kms and a 2nd or 3rd-order polynomial baseline was subtracted.
The N10 special-purpose baseline fitting and extra B-spline baseline removal procedures were also
performed on the data of the second survey.  The final spectra were cubic spline interpolated onto the
final velocity scale of the MS-tip data.
Finally, the spectra were resampled onto the Galactic coordinate grid and combined with the MS-tip data
with exposure time weighting.

\section{Results}
\label{sec:results}

Previous, higher-resolution interferometry data show a regularly rotating inner \hi disk extending out to
$\sim$4\arcmin~in radius \citep{Cohen79,Shostak89,Wilcots98,Manthey08} with a 
significant inclination (Shostak \& Skillman derive $\sim$40\dgr while Wilcots \& Miller
find a larger value of 60--90\degr) and rotation velocity of $\sim$30 \kmse.
Several extended \hi features have been seen with kinematics consistent with either
a warped disk \citep{Cohen79} or a counter-rotating disk \citep{Wilcots98}.  There are
several ``spurs,'' a southern ``streamer'' (extending $\sim$30\arcmin~to the south),
and a small gas cloud to the NE \citep[][$\sim$30\arcmin~to the northeast]{Manthey08}.

\begin{figure*}[ht!]
\begin{center}
\includegraphics[angle=0,scale=0.60]{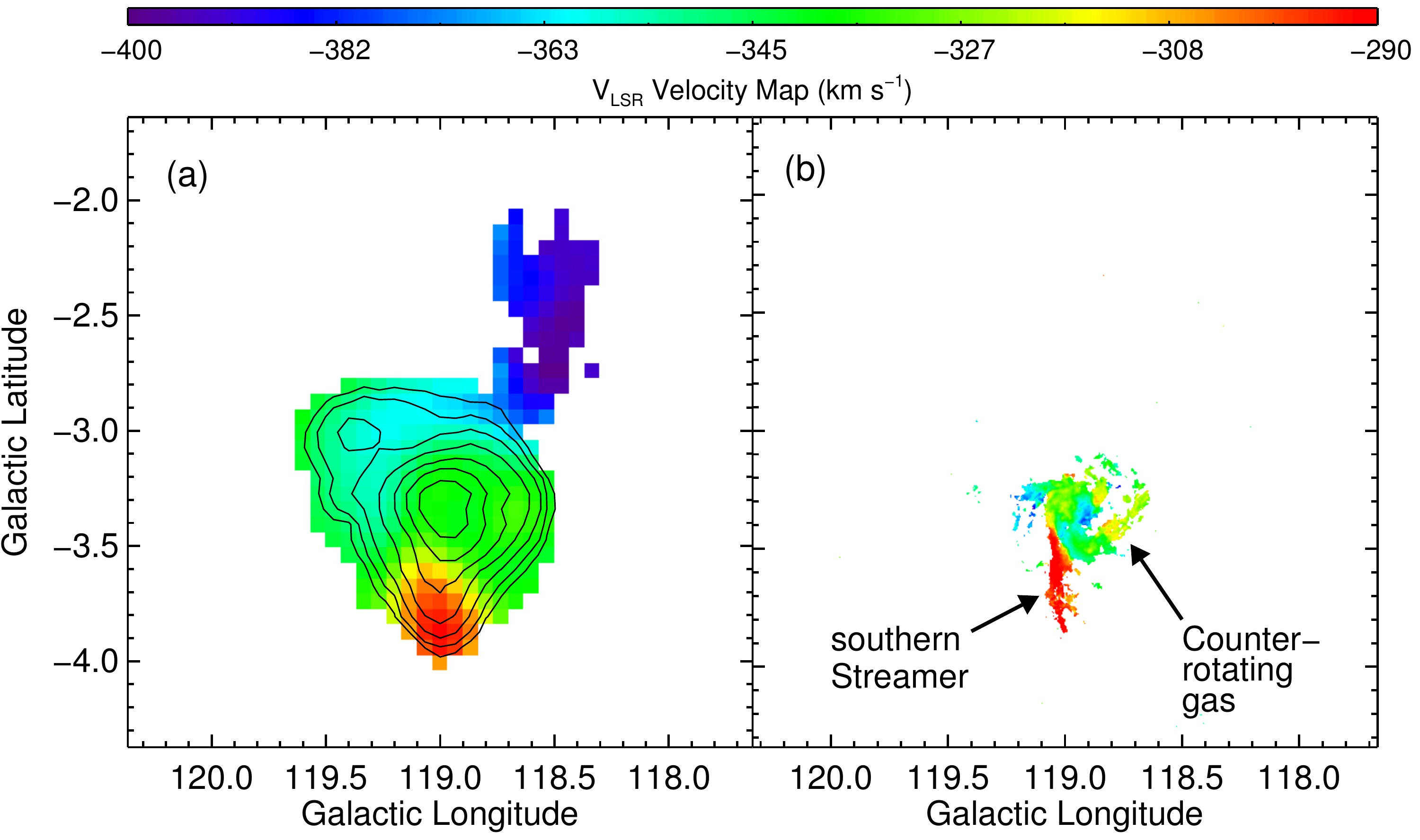}
\end{center}
\caption{{\rm (a)} GBT velocity map of IC~10 and the new northern extension.  Pixels with no IC~10 associated
emission are masked out.  Column density contours are the same as in Figure \ref{fig_coldens_stream}a.
{\rm (a)} IC~10 velocity map of the \citet{Manthey08} WSRT data.  The southern streamer and ring of counter-rotating
gas are visible.  The northern extension is not detected in the WSRT data.}
\label{fig_velmap}
\end{figure*}

The Figure \ref{fig_coldens}a color image shows the GBT column density map of the entire IC~10
($-419.0<$\vlsre$<-251.5$ \kmse)\footnote{Note that at the position of IC~10 the Galactic
and equatorial coordinate axes are nearly aligned.} with a 3$\sigma$ sensitivity of
$\sim$1.9$\times$10$^{18}$ atoms cm$^{-2}$, while contours show column density of the \citet{Manthey08} WSRT
data for comparison.
The overall high-resolution morphology is also seen in our GBT data, albeit at lower spatial resolution than
in the interferometry data.\footnote{A detailed comparison of LITTLE THINGS interferometry data and the GBT data
will be presented in a future paper (T.\ Ashley et al., in preparation).}
The southern ``streamer'' stretches $\sim$37\arcmin\ southward of the IC~10 center
(Fig.\ \ref{fig_coldens}a) and reaches the highest velocities in IC~10 at \vlsr$\approx$$-280$ \kms
(Fig.\ \ref{fig_coldens}b) .
The NE cloud stretches $\sim$37\arcmin\ to the northeast but to the lowest velocities
in the main body of IC~10 at \vlsr$\approx$$-380$ \kmse.

The high sensitivity and low side-bands of the GBT allows us to detect faint \hi emission that was
not seen (detectable) in previous datasets.  Figure \ref{fig_coldens_stream}a shows a smoothed
(with 12\arcmin~filter, giving an effective $\sim$15\arcmin~resolution) column density map integrated
over velocities $-412.5.0<$\vlsre$<-388.4$ \kms
with a 3$\sigma$ sensitivity of $\sim$4.0$\times$10$^{17}$ atoms cm$^{-2}$.
This figure reveals a newly-found \hi feature extending to the northwest (orientation of $\sim$25\dgr
west of north), hereafter referred to as the ``extension''.

Figure \ref{fig_coldens_stream}b shows the latitude--velocity diagram for the longitude
range where the \hi extension is prominently seen (118.40$<$$l$$<$118.67\degr).  Contours show the emission
from the entire body of IC~10 as shown in Figure \ref{fig_coldens}a.  The feature is seen as an arc extending
$\sim$1.3\dgr to the north ($b$$\sim$$-2.1$\degr) and to \vlsr$\approx$$-410$ \kms ($\sim$65 \kms below
the IC~10 systemic velocity).  At a distance of 805 kpc this corresponds to a projected
length of $\sim$18.3 kpc.  The average width (in $l$) of the extension is $\sim$0.37\dgr or 5.2 kpc
(at 805 kpc).  There is also a higher-density concentration at the end of the feature (as seen in both
panels of Fig.\ \ref{fig_coldens_stream}).  It is difficult to discern the extension from the rest of
IC~10 near the center of the galaxy, but the extension appears to stretch from near the center of IC~10
to its outskirts.  The velocity map (Fig.\ \ref{fig_velmap}a) shows a smooth velocity gradient from the
extension to the edge of IC~10 and continuing into the main body.
For the latitudes where the extension is clearly seen we summed spectra across the extension (in longitude)
and fit the resulting velocity profiles with a Gaussian.
The column density of the extension drops with distance from the center of IC~10 (as also seen in
Fig.\ \ref{fig_coldens_stream}a) but increases near the end.
The mean column density of the extension is $\sim$7$\times$10$^{17}$ atoms cm$^{-2}$.
The mass of the extension is $\sim$7.1$\times$10$^5$ ($d$/805 kpc)$^2$ \msun which is only $\sim$0.75\%
of the 9.5$\times$10$^7$ ($d$/805 kpc)$^2$ \msun mass of the entire IC~10 galaxy as measured by our
GBT data.

In the latitude--velocity diagram
(Fig.\ \ref{fig_coldens_stream}b) the southern streamer and extension appear symmetric (in position and velocity)
about the IC~10 center.  They are also nearly aligned on the sky (although having different angular
lengths) as seen in Figure \ref{fig_coldens_stream}a and the WSRT data in Figure \ref{fig_coldens}b.
However, they have column densities that differ by a factor of $\sim$70$\times$ which makes a connection
between these two features tenuous.

\begin{figure*}[ht!]
\includegraphics[angle=0,scale=0.58]{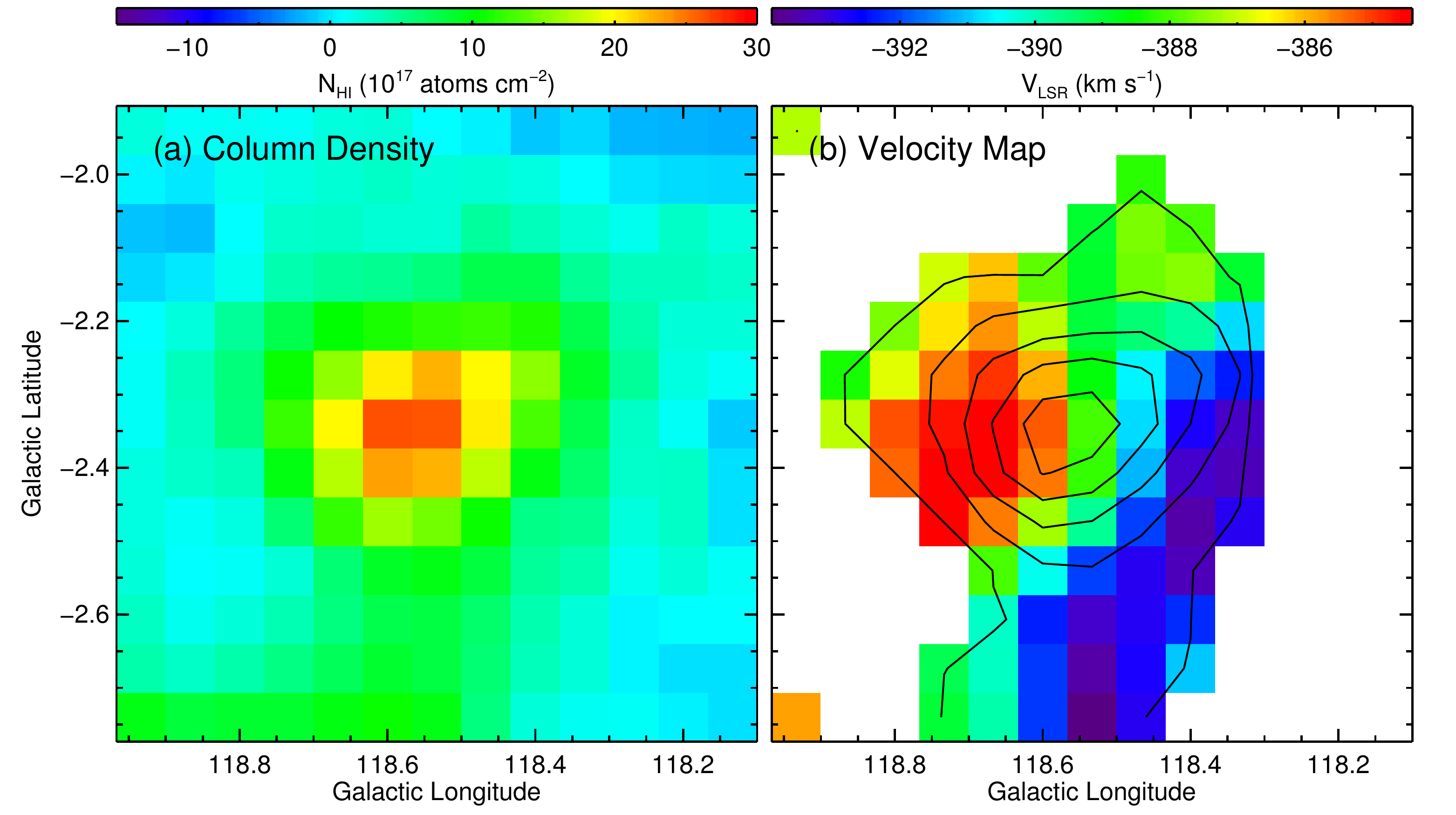}
\caption{({\em a}) Column density map of the high density region at the end of the northern
extension (in units of 10$^{17}$ atoms cm$^{-2}$).
({\em b}) Velocity map (\vlsre) for region in ({\em a}) with column density contours
(at levels of \nhie=12--28$\times$10$^{17}$ atoms cm$^{-2}$ in steps of 4$\times$10$^{17}$).}
\label{fig_blob}
\end{figure*}

The higher column density region at the end of the extension is azimuthally-symmetric
(Fig.\ \ref{fig_blob}a), $\sim$26\arcmin\ ($\sim$6.1 kpc) across, and has a mass of $\sim$3.5$\times$10$^5$
($d$/805 kpc)$^2$ \msune.  The velocity map in Figure \ref{fig_blob}b shows a coherent velocity gradient
across the region ($\sim$10 \kms from one end to the other, or $\sim$1.7 \kms kpc$^{-1}$) that is
perpendicular to the length of the extension (which has a velocity gradient along it).

\section{Origin of the \hi Extension}
\label{sec:origin}

There are several possible origins of the newly-found northern extension: interactions of IC~10
with M31 (either tidal or ram pressure), interaction of IC~10 with the diffuse intergalactic medium (IGM),
interactions of IC~10 with a companion galaxy, cold accretion, or stellar feedback (a suggested contributing
factor in the creation of the Magellanic Stream; Nidever et al.\ 2008) from the IC~10 starburst itself.

Unlike most LG dwarf galaxies outside the MW, IC~10 has a measured proper
motion from VLBA masers \citep{Brunthaler07}.  This allows us to produce a fairly accurate model of the
IC~10 orbit.  For IC~10 we use our measured GBT radial velocity (\vlsr$=-338.5$ \kmse),
the Brunthaler et al.\ proper motions, and a distance of 805 kpc \citep{Sanna08}.
For M31, we use a distance of 770 kpc \citep{vanderMarel08}, a radial velocity of $-301$ \kms
\citep{Courteau99}, and the recent $HST$ proper motions from \citet{vanderMarel12}.  A mass of
1.4$\times$10$^{12}$ \msun
is used for M31 \citep{Watkins10}\footnote{The uncertainties in the M31 mass have little impact on the
final IC~10 orbit.}, modeled as a static Plummer potential with 9 kpc softening parameter,
while a 3-component static MW potential is used \citep{Johnston95}.  We adopted 
$R_0$=8.29 for the solar radius and $V_0$=239 \kms for the local standard of rest velocity \citep{McMillan11},
as well as the \citet{Schoenrich10} values for the sun's peculiar velocity.
A modified version of the \citet{Hut03}
Gravitylab N-body integrator code was used to perform the orbit calculations.  A Monte Carlo simulation
with 1000 mocks sampling the error space for all input parameters was performed and used to calculate
the 1$\sigma$ uncertainties in the orbit (dash-dotted lines in Fig.\ \ref{fig_coldens_stream}).

As seen in Figure \ref{fig_coldens_stream}a the orbit of IC~10 runs diagonally from the southeast to the
northwest.   While the orientation of the extension is generally aligned with the orbit, it is {\em leading}
IC~10 in its orbit.  Therefore, the new extension cannot be due to ram pressure from the M31 halo gas
or the intergalactic medium because ram pressure produces {\em trailing} tails.  Moreover, the
latitude--velocity diagram in Figure \ref{fig_coldens_stream}b shows that the change in velocity of the
orbit is very small
and cannot reproduce the large velocity gradient along the extension.  This is in stark contrast to the
Magellanic Stream where the large observed radial velocity gradient is also seen in the LMC's orbit
\citep[e.g.,][]{Besla10} indicating that the Milky Way's gravitational force has been an important factor in
shaping the velocities of the Stream.  In addition, IC~10's orbital period about M31 is 6.3$^{+4.5}_{-0.73}$ Gyr
(with no mock periods shorter than 3.8 Gyr and $\sim$13\% larger than a Hubble time)
with the last perigalacticon 1.88$^{+0.34}_{-0.04}$ Gyr ago at a distance of 82$^{+94}_{-26}$ kpc.
It is, therefore, very unlikely that the newly-found IC~10 \hi extension can be explained by an M31--IC~10
tidal interaction.

Could stellar feedback from the starburst have created the \hi extension?  According to \citet{Massey07},
the high number of Wolf-Rayet stars and lack of young red supergiants indicates that IC~10's starburst
is quite young, roughly $\sim$10 Myr.  Therefore, to create an $\sim$18 kpc--long extension in 10 Myr would
require an outflow velocity (in the plane of the sky) of 18 kpc/0.01 Gyr$\approx$1800 \kmse!  This is
vastly larger than the extension velocity offset of $\sim$65 \kms observed.  Due to this timing argument
we rule out a starburst origin of the extension.

Nearby primordial gas being accreted onto IC~10 counter to the rotation of the main body could also
explain these outer \hi features.  Cold accretion has been suggested as a possible trigger for isolated
starburst dwarf galaxies.  The combined mass of the counter-rotating \hi in the outskirts of IC~10
(beyond 13\arcmin~from the center) and the \hi extension from our data is $\sim$3.3$\times$10$^7$ \msune.
However, there is no observational evidence of a population of $\gtrsim$10$^7$ \msun 
\hi clouds in intergalactic space that are fueling star formation in galaxies \citep{Sancisi08}, although
smaller clouds or filaments could exist in the IGM and be accreted.
Therefore, if the counter-rotating gas and extension are from cold accretion alone, then the gas filament
would be fairly massive.
The detection of a stellar component would rule out a primordial gas cloud origin.

This leaves us with an interaction origin.  If there was an interaction, then where is the companion
galaxy?  The higher column density region at the end of the \hi extension with a coherent velocity gradient
is a potential candidate for an interacting dwarf galaxy.  If the velocity gradient is due to rotation,
then an enclosed mass of 1.74$\times$10$^7$ \msun is required to sustain 5 \kms circular orbits at a
radius of 3 kpc at the edge of the region (not accounting for any inclination effects).
This would imply a massive dark matter or stellar component.
We searched the publicly available
2MASS \citep{Skrutskie06}, WISE \citep{Wright10}, and POSS \citep{Minkowski63} images for stellar components
of both the extension and the higher density region but found no convincing detection.  This close to the MW
midplane ($b$=$-3.3$\degr) the dust extinction is high making a faint stellar component difficult to
detect.  Moreover, if the companion galaxy is similar to Leo T \citep{Irwin07}, which contains
$\sim$3$\times10^5$ \msun of \hi and at a distance of $\sim$420 kpc was difficult to detect in SDSS
star counts, then we would likely not have seen such a low surface brightness feature in the
``shallow'' images we inspected.
Deep and wide-field photometry, like those obtained by \citet{Sanna10}, would
likely be necessary to properly probe any stellar features.
Another possibility is that the companion galaxy is located somewhere else and has so far eluded detection.

An interaction origin would naturally explain the starburst
(i.e., as triggered by the interaction) and possibly the counter-rotating gas in the outer regions of IC~10
(i.e., by accreted gas from the companion galaxy).
Moreover, an interaction origin of the IC~10 starburst would answer a long-standing mystery of how such a small
and apparently isolated galaxy could have such a high star formation rate.  Therefore, we find that a scenario of
a recent dwarf-dwarf interaction is the most likely explanation for the newly-found \hi extension.

The present discovery is reminiscent of another BCD, namely NGC 4449.  NGC 4449 was proposed to be
interacting with a nearby dwarf galaxy \citep{Theis01}, creating disturbed kinematics and morphology.
However, upon further investigation a small, tidally distorted dwarf spheroidal was found nearby
\citep{Rich12,delgado12}.  This new companion to NGC 4449 would not had been found without deep
optical observations.  Another example is the starburst galaxy IZw18 that was recently found to have a
long \hi tail likely produced by an interaction \citep{Lelli12}.
Therefore, some apparently isolated dwarf galaxies may not be isolated at all,
but have their bursts of star formation possibly triggered by interactions with undiscovered companions.

\acknowledgements

We thank Eric Wilcots, Roeland van der Marel, and Antonela Monachesi for useful discussions.
D.L.N. was supported by a Dean B.\ McLaughlin fellowship
at the University of Michigan.
C.T.S. and E.F.B. acknowledge support from NSF grant AST 1008342.
This work was funded in part by the National Science Foundation
through grant AST-0707468 to C.E.S.
The National Radio Astronomy Observatory is operated by Associated Universities, Inc., 
under cooperative agreement with the National Science Foundation.

\end{document}